\documentclass[prb,reprint]{revtex4-2}
\usepackage{graphicx,amssymb,amsfonts,amsmath}
\usepackage{color}
\usepackage{mathtools}
\usepackage{latexsym}
\usepackage{graphicx}
\usepackage{float}
\usepackage{dcolumn}
\usepackage{bm}
\usepackage{amssymb}
\usepackage{amsmath}
\usepackage{mathtools}
\usepackage[space]{grffile}
\usepackage{xcolor}
\usepackage{comment}
\usepackage{hyperref}
\hypersetup{
    colorlinks=true,
    linkcolor=blue,
    filecolor=blue,      
    urlcolor=blue,
    citecolor=blue
    }

\begin{document}

\title{Universal Model of Floquet  Operator Krylov Space}
\author{Hsiu-Chung Yeh}
\author{Aditi Mitra}
\affiliation{
Center for Quantum Phenomena, Department of Physics,
New York University, 726 Broadway, New York, New York, 10003, USA
}

\begin{abstract}
It is shown that the stroboscopic time-evolution under a Floquet unitary, in any spatial dimension, and of any Hermitian operator,  can be mapped to an operator Krylov space which is identical to that generated by the edge operator of the non-interacting Floquet transverse-field Ising model (TFIM) in one-spatial dimension, and with inhomogeneous Ising and transverse field couplings. The  latter has four topological phases reflected by the absence (topologically trivial)
or presence (topologically non-trivial) of edge modes at $0$ and/or $\pi$ quasi-energies. It is shown that the Floquet dynamics share certain universal features characterized by how the Krylov
parameters vary in the topological phase diagram of the Floquet TFIM
with homogeneous couplings. These results are highlighted through examples, all chosen for numerical convenience to be in one spatial dimension: non-integrable Floquet spin $1/2$ chains and Floquet $Z_3$ clock model where the latter hosts period-tripled edge modes.   
\end{abstract}
\maketitle

\section{Introduction}
The nonequilibrium dynamics of interacting quantum systems is a challenging problem as few non-perturbative analytic methods exist, and exact numerical simulations are restricted to small system sizes. Recently, the study of operator dynamics in Krylov space \cite{Recbook,parker2019universal,rabinovici2021operator,balasubramanian2022quantum,caputa2022quantum,liu2023krylov}, has emerged as a powerful method because this approach maps an interacting problem to a non-interacting one, making it amenable to a host of methods valid for single particle quantum mechanics \cite{Yates20,Yates20a,yeh2023slowly}. However, this approach has been primarily applied to continuous time evolution,
where the operator Krylov space corresponds to a single particle hopping  on a tight-binding lattice with nearest-neighbor and inhomogeneous hopping.  Extending Krylov methods to Floquet systems \cite{Yates21,Yates22,suchsland2023krylov} is particularly important as Floquet dynamics is qualitatively different from continuous time dynamics because energy is conserved in the former only up to integer multiples of the driving frequency \cite{SondhiRev,OkaRev}, resulting in  spectra that are $2\pi$ periodic.  This periodicity is behind novel phenomena such as new topological phases with no counterpart in continuous time evolution \cite{jiang2011majorana,Rudner13,Potter16,asboth2014chiral,Khemani16,Else16b,Roy16,Po16,Potter17,Morimoto17,Fidk19,Yates19}. 

In this work we derive a remarkably simple operator Krylov space picture under stroboscopic (Floquet)  time evolution. We show that the Krylov space of any Hermitian operator undergoing unitary Floquet dynamics, independent of the spatial dimension,  is identical to the Krylov space of a 
non-interacting problem in one dimension. The latter is the Floquet transverse field Ising model with inhomogeneous couplings (ITFIM).  For homogeneous couplings, the Floquet transverse field Ising model (TFIM) is well documented as having four topological phases \cite{Sen13,Khemani16,Yates19,yeh2023decay}: a trivial phase with no edge modes, a $0$ ($\pi$)-mode phase with an edge mode at zero ($\pi$) quasi-energy, and a $0$-$\pi$ phase where both $0$ and $\pi$ edge modes exist together. We show that all operator dynamics in Krylov space can be mapped to the edge mode dynamics of a suitable  ITFIM. In addition, the dynamics has certain universal features reflected in the trajectory that the inhomogeneous couplings take in the topological phase diagram of the TFIM, on moving from the edge to the bulk of the Krylov chain. This allows us to capture the exact dynamics quite faithfully, at least at short and intermediate times, by employing a much smaller number of Krylov parameters than the size of the Hilbert space.  

The paper is organized as follows. In Section \ref{Sec2} we describe the operator Krylov space of a generic Hermitian operator undergoing stroboscopic time-evolution due to a generic unitary, showing that the Krylov space can be parameterized by angles, which we dub the Krylov angles. In Section \ref{Sec3} we show that a Krylov space with the very same Krylov angles is generated by the edge operator of a ITFIM, under Floquet time evolution by the ITFIM, where the couplings  of the ITFIM correspond to the Krylov angles. 
In Section \ref{Sec4} we present applications of this mapping. In particular we show that this mapping can provide physically motivated truncation schemes for the dynamics of quasi-conserved operators as they manifest as topologically protected edge modes of the ITFIM in Krylov space. Although our examples of quasi-conserved operators are edge operators of several models, the truncation scheme is applicable 
for more general conserved operators with applications to integrable models, weakly non-integrable models, and scar
states in more generic non-integrable models. In Section \ref{Sec5} we present our conclusions, and details are relegated to six appendices. 

\section{Operator Krylov Space from a Floquet Unitary}\label{Sec2}
In order to construct the Floquet Krylov space for a given operator, one follows the Arnoldi iteration \cite{arnoldi1951principle}. One begins with an initial operator $O_0$ and then generates new operators via the Floquet unitary $U$. Finally, a set of orthonormal operators $\{ \mathcal{O}_i \}$ are generated by the Gram-Schmidt procedure. Explicitly, one initially sets $\mathcal{O}_0 = O_0$ and performs the algorithm below for an integer $i>0$
\begin{align}
    &|\mathcal{O}_i^\prime) = K|\mathcal{O}_{i-1}) - \sum_{j=1}^{i-1}|\mathcal{O}_j)(\mathcal{O}_j|K|\mathcal{O}_{i-1});\\
    &|\mathcal{O}_i) = \frac{|\mathcal{O}_i^\prime)}{\sqrt{(\mathcal{O}_i^\prime|\mathcal{O}_i^\prime)}},
\end{align}
where $|\mathcal{O}_i)$ is the vector representation of $\mathcal{O}_i$ and $K|\mathcal{O}_i) = U^\dagger \mathcal{O}_i U$. The inner product of two operators is defined as $(A|B) = \text{Tr}[A^\dagger B]/N$, where $A$ and $B$ are $N \times N$ matrices, with $N$ being the size of Hilbert space. $K$ has an upper-Hessenberg form $(\mathcal{O}_i|K|\mathcal{O}_j)=0$ for $i > j+1$ \cite{Yates21,suchsland2023krylov}. In addition, Ref.~\cite{suchsland2023krylov} showed that $K$ has the following simple structure parameterized by $\{ a_i \}$, $\{ b_i \}$ and $\{ c_i \}$  in the orthonormal bases $\{ |\mathcal{O}_i) \}$:
\begin{align}
&b_i=(\mathcal{O}_i|K|\mathcal{O}_{i-1});
&\frac{a_i}{c_i}c_j=(\mathcal{O}_i|K|\mathcal{O}_j)\ \text{for}\ i \leq j,
\label{Eq: K matrix element}
\end{align}
where $a_0 = c_0$. For example, $K$ has the following explicit expression for $N = 6$,
\begin{align}
    K=\begin{pmatrix}
        a_0 & c_1 & c_2 & c_3 & c_4 & c_5\\
        b_1 & a_1 & \frac{a_1}{c_1}c_2 & \frac{a_1}{c_1}c_3 & \frac{a_1}{c_1}c_4 & \frac{a_1}{c_1}c_5\\
        0 & b_2 & a_2 & \frac{a_2}{c_2}c_3 & \frac{a_2}{c_2}c_4 & \frac{a_2}{c_2}c_5\\
        0 & 0 & b_3 & a_3 & \frac{a_3}{c_3}c_4 & \frac{a_3}{c_3}c_5\\
        0 & 0 & 0 & b_4 & a_4 & \frac{a_4}{c_4}c_5\\
        0 & 0 & 0 & 0 & b_5 & a_5
    \end{pmatrix}.\label{K6}
\end{align}
If the initial operator $O_0$ is Hermitian, the orthonormal operators $\{ \mathcal{O}_i \}$ are also Hermitian which implies $\{ a_i \}$, $\{ b_i \}$ and $\{ c_i \}$ are all real numbers. With the condition of unitarity, $K$ is an orthogonal matrix, $K^\intercal K = \mathbb{I}$, and $\{ a_i \}$, $\{ b_i \}$, $\{ c_i \}$ are not independent variables. In particular, we show that the following relations hold between the matrix elements, see Appendix \ref{Sec:A}.
\begin{align}
    &a_i = \cos\theta_i\cos\theta_{i+1};
    &b_i = \sin\theta_i;\nonumber\\
    &c_i = (-1)^i\cos\theta_{i+1} \prod_{j=1}^i\sin\theta_j,
    \label{Eq: Krylov abc}
\end{align}
where $\theta_0 = \theta_{N} = 0$ and $\theta_i \in [0,\pi]$. This is consistent with the numerical observation in \cite{suchsland2023krylov} that $|a_i|$, $|b_i|$, $|c_i| \leq 1$. Essentially, all the information of the Floquet Krylov space is encoded in the Krylov angles $\{ \theta_i \}$, which can be derived from computing only the   $\{ a_i \}$.

\section{A Universal Model} \label{Sec3}
Identical matrix elements of $K$ in  Krylov space, or equivalently $\{ a_i \}$, $\{ b_i \}$ and $\{ c_i \}$ in \eqref{Eq: Krylov abc},  can be generated by the ITFIM with open boundary conditions
\begin{align}
&U_{\text{ITFIM}} = U_z U_{xx},
\end{align}
with
\begin{align}
U_z = \prod_{j=1}^{N/2} e^{-i\frac{\theta_{2j-1}}{2}\sigma_j^z}; U_{xx}=\prod_{j=1}^{(N-2)/2} e^{ -i\frac{\theta_{2j}}{2}\sigma_j^x \sigma_{j+1}^x},
\end{align}
where we assume $N$ is even, and the boundaries are at $j=1,N/2$. The Krylov angle $\theta_{2j-1}$ ($\theta_{2j}$) corresponds to the local transverse-field (Ising coupling) of the model. This is a non-interacting system since $U_{\text{ITFIM}}$ is bilinear in Majorana fermions $\gamma_{\ell}, \ell \in 1,\ldots,N$.
By constructing the Krylov space of the edge Majorana $\gamma_1$ undergoing Floquet evolution due to $U_{\text{ITFIM}}$, the $K_{\text{ITFIM}}$ in the Krylov space orthonormal basis has the exact same matrix elements as the $K$ in \eqref{Eq: K matrix element} and \eqref{Eq: Krylov abc}, see details in Appendix \ref{Sec:B}. We emphasize that this mapping is always valid regardless of the spatial dimension, or local Hilbert space of the original model. 
Therefore, the study of the autocorrelation function of some Hermitian operator $O_0$ in any Floquet model is equivalent to the study of the autocorrelation function of the edge Majorana $ \gamma_1$, in the non-interacting Floquet ITFIM with local couplings $\{ \theta_i \}$. Since this mapping is at the level of operator evolution, the equivalence also extends to objects beyond autocorrelation functions, such as operator entanglement.

\begin{figure*}
    \includegraphics[width=0.24\textwidth]{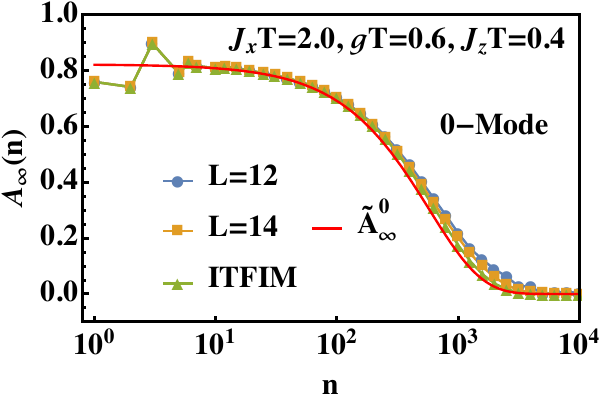}
    \includegraphics[width=0.24\textwidth]{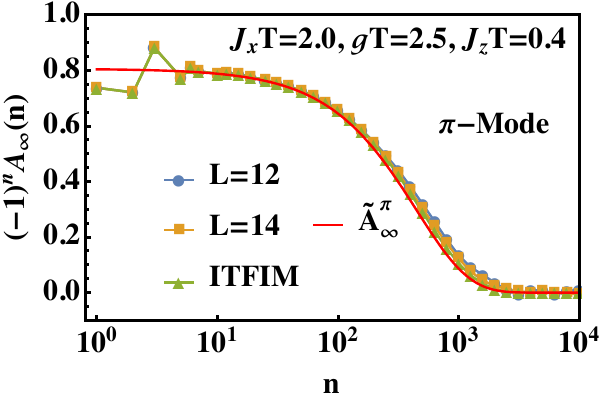}
    \includegraphics[width=0.24\textwidth]{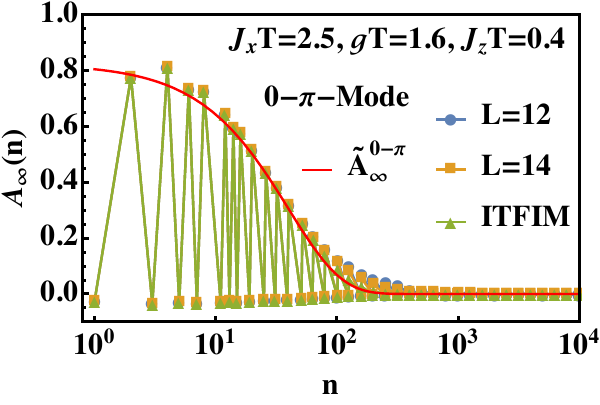}
    \includegraphics[width=0.24\textwidth]{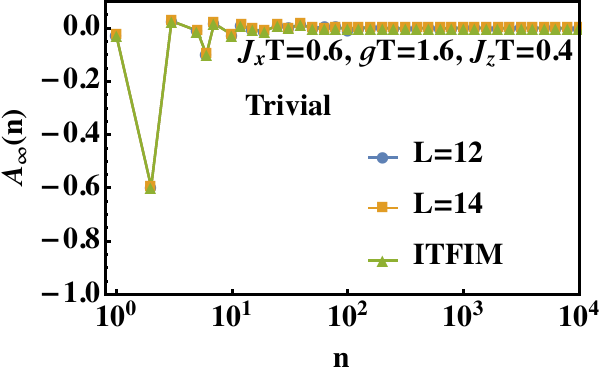}
    \includegraphics[width=0.24\textwidth]{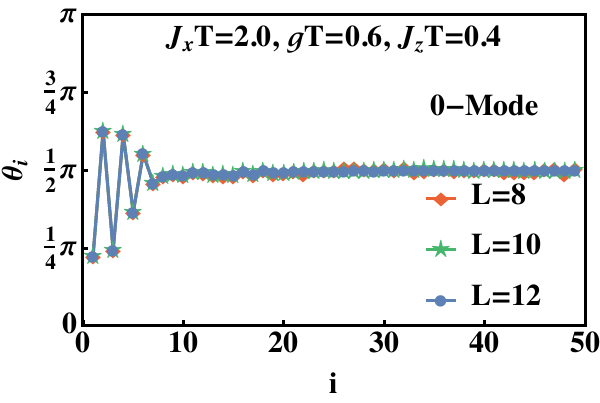}
    \includegraphics[width=0.24\textwidth]{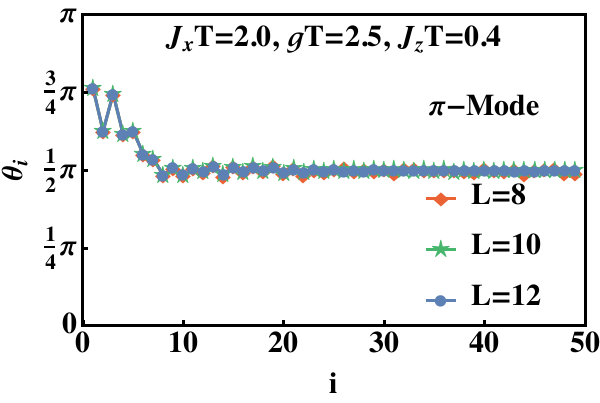}
    \includegraphics[width=0.24\textwidth]{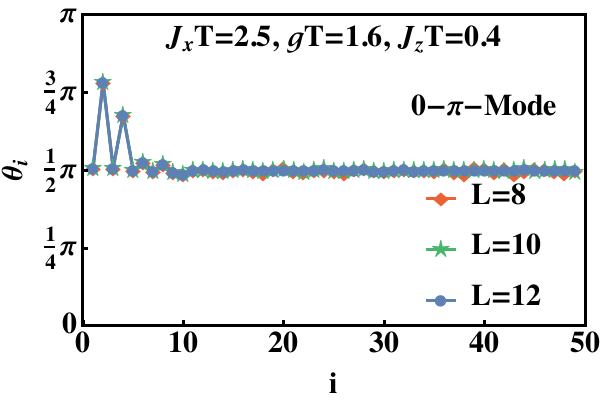}
    \includegraphics[width=0.24\textwidth]{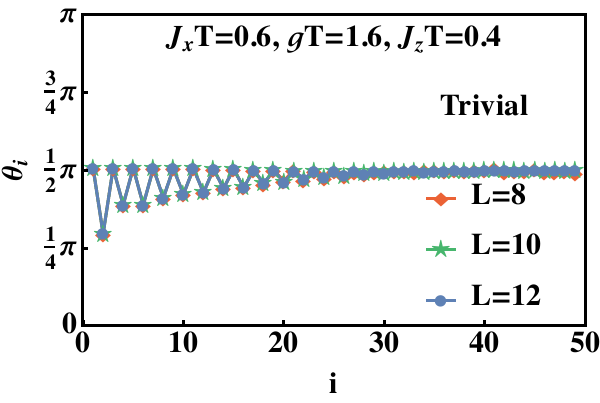}

    \caption{Top panels: Infinite temperature autocorrelation function of the edge spin $\sigma_1^x$ for the unitary \eqref{Eq: Perturbed TFIM} in four different topological phases, $0$-mode, $\pi$-mode, $0$-$\pi$-mode and trivial from left to right, and from ED for $L=12,14$. Bottom panels: Krylov angles for the same, for $L=8,10,12$. $K_{\text{ITFIM}}$ is truncated to a $50 \times 50$ matrix and constructed from the Krylov angles for $L=12$ with $\theta_{50}=\pi/2$. The corresponding  autocorrelation  obtained as a return amplitude, \eqref{Eq: itfimR} (green lines in top panels), and from further approximations, \eqref{Eq: Approx A} (red lines in the top panels) show good agreement with ED. This supports the physical picture that the long-lived autocorrelation is an approximate edge mode of the ITFIM, and that the survival of the edge mode is related to the segment in Krylov space where the angle $\theta_i$ is confined within the relevant topological phase.}
    \label{Fig: perturbed TFIM}
\end{figure*}

While $K$ in \eqref{K6} is non-local and the right way to truncate it is not obvious, the fact that it
is generated by $U_{\rm ITFIM}$ with local couplings denoted by the Krylov angles $\theta_i$, allows us to develop physically motivated guidelines for performing truncations based on the
behavior of these Krylov angles. To demonstrate this, we study the following models, all chosen in one-dimension for numerical convenience. The first model corresponds to an open spin $1/2$ chain with $L$ sites:
\begin{align}
    &U_1 = e^{-i\frac{T}{2}J_z H_{zz}}e^{-i\frac{T}{2}g H_z}e^{-i\frac{T}{2}J_x H_{xx}}; \label{Eq: Perturbed TFIM}
\end{align}
where
\begin{align}
    H_{xx} = \sum_{i=1}^{L-1} \sigma_i^x\sigma_{i+1}^x
    ;H_{zz} = \sum_{i=1}^{L-1} \sigma_i^z\sigma_{i+1}^z;H_{z} = \sum_{i=1}^{L} \sigma_i^z.
\end{align}
$U_1$ is the Floquet TFIM perturbed by a transverse Ising interaction $J_z\neq 0$. The second model is the generalization of the TFIM to the $Z_3$ clock model \cite{sreejith2016parafermion}
\begin{align}
    U_2 = e^{-i\frac{T}{2}H_h}e^{-i\frac{T}{2}H_J},
    \label{Eq: Z3 clock model}
\end{align}
where
\begin{align}
    &H_J = J\sum_{i=1}^{ L-1} \sigma_i\sigma_{i+1}^\dagger + \text{H.c.}; &H_h = h\sum_{i=1}^L \tau_i + \text{H.c.},\\
    &\sigma = \begin{pmatrix}
        1 & 0 & 0\\
        0 & e^{i2\pi/3} & 0\\
        0 & 0 & e^{i4\pi/3}
    \end{pmatrix};
    &\tau = \begin{pmatrix}
        0 & 0 & 1\\
        1 & 0 & 0\\
        0 & 1 & 0
    \end{pmatrix}.
\end{align}
We study the above models with open boundary conditions because these models host long lived edge modes, and hence have non-trivial autocorrelation functions. 
To further support the applicability of ITFIM, we also study the autocorrelation function of an edge and a bulk spin for the Floquet transverse-field XYZ model in Appendix \ref{Sec:D}. Note that the Hilbert spaces of the model we are studying and the effective model to which the Krylov dynamics maps, the ITFIM, may be completely different as highlighted from the example of the $Z_3$ clock model. 

\section{Results and Discussion} \label{Sec4}
\begin{figure}[h!]
    \centering
    \includegraphics[width=0.3\textwidth]{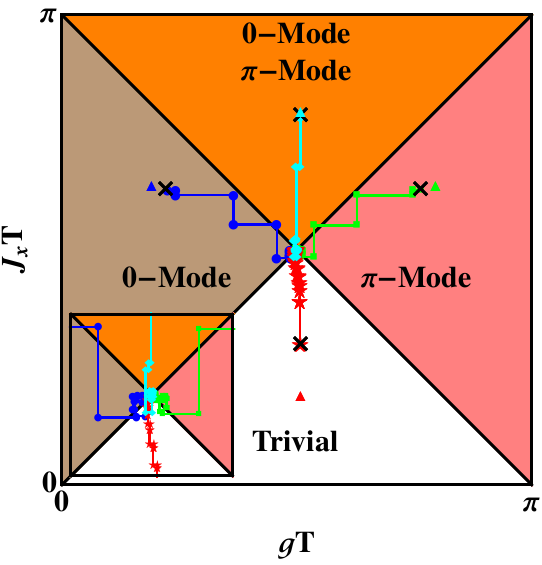}
    
    \caption{The trajectory of local couplings from the edge to the bulk for the ITFIM, where $\theta_{2j-1}$ ($\theta_{2j}$) corresponds to the local transverse-field $g_jT$ (Ising coupling $J_{xj}T$). The Krylov angles are for $L=12$, for model \eqref{Eq: Perturbed TFIM}, with the black crosses
    denoting the local couplings on the first Krylov site. The four  triangles mark the non-perturbed limit  ($J_z=0$) of the top panels in Fig.~\ref{Fig: perturbed TFIM}. The perturbation leads to inhomogeneous couplings. The local couplings on the first site take values close to the unperturbed limit. In contrast, in the bulk, the local couplings approach $\pi/2$ (bottom panels in Fig.~\ref{Fig: perturbed TFIM}).  The phase diagram shows that the autocorrelation will eventually decay since the couplings in the bulk cannot allow for the existence of edge modes.}
    \label{Fig: Phase diagram}
\end{figure}

We compute the infinite temperature autocorrelation function, defined below, by exact diagonalization (ED),
\begin{align}
    A_\infty(n) = \frac{1}{N}\text{Tr}[O_0(n)O_0].
\end{align}
We simulate equally spaced data points on a logarithmic time scale to capture features in the long time scale. For $U_1$ \eqref{Eq: Perturbed TFIM}, we take the seed operator to be the edge operator $O_0 = \sigma_1^x$  (Fig.~\ref{Fig: perturbed TFIM}) and we take $gT,J_xT$ in \eqref{Eq: Perturbed TFIM} to lie within the four different topological phases of the Floquet TFIM, (triangles in Fig.~\ref{Fig: Phase diagram}). These are phases that host a $0$-mode ($J_xT = 2.0$ and $gT = 0.6$), a $\pi$-mode ($J_xT = 2.0$ and $gT = 2.5$), $0$-$\pi$-modes ($J_xT = 2.5$ and $gT = 1.6$) and a trivial phase with no edge modes ($J_xT = 0.6$ and $gT = 1.6$). We take the strength of the integrability breaking perturbation to be $J_zT = 0.4$. Due to the perturbation, the edge modes are almost strong modes \cite{yeh2023decay,Yates19,Yates21,Fendley17,Nayak17,Yao20} where the autocorrelation decays to zero at a system size independent, albeit long time (top panels of Fig.~\ref{Fig: perturbed TFIM}).   The Krylov angles of these four setups are shown in the lower panels of Fig.~\ref{Fig: perturbed TFIM}, where the angles approach $\pi/2$ in the bulk of the Krylov chain. Since the full Hilbert space size is $N = 2^L$, it is impractical to exhaust all Krylov angles numerically. Therefore, we truncate $K_{\text{ITFIM}}$ into a $50 \times 50$ matrix and construct its matrix elements from the Krylov angles for $L=12$. We set the boundary condition of the truncation as $\theta_0 = 0$ and $\theta_{50} = \pi/2$. Due to the truncation, $K_{\text{ITFIM}}^\intercal K_{\text{ITFIM}}$ is not equal to identity. Instead, it is a diagonal matrix in Krylov space with matrix elements $(1, \ldots, 1, 0)$ along the diagonal. The autocorrelation can also be computed from the $K_{\text{ITFIM}}$ from the relation
\begin{align}
    A_\infty(n) = (K_{\text{ITFIM}}^n)_{1,1}.\label{Eq: itfimR}
\end{align}
Above $(K_{\text{ITFIM}}^n)_{1,1}$ is the upper left-most matrix element of $K_{\text{ITFIM}}^n$.
\eqref{Eq: itfimR}  encodes the equivalence between the autocorrelation in ED and the return amplitude of the edge Majorana of the ITFIM after $n$ periods. Here we compute \eqref{Eq: itfimR} in the single Majorana basis because ITFIM is bilinear in Majorana fermions making $K_{\rm ITFIM}$ a sparse matrix in this basis, see details in Appendix \ref{Sec:B}. Fig.~\ref{Fig: perturbed TFIM} shows that $K_{\text{ITFIM}}$ captures the initial decay of the autocorrelation from ED, while it decays to zero somewhat faster than ED due to the truncation of $K_{\text{ITFIM}}$, see Appendix \ref{Sec:C}.

\begin{figure*}
    \centering
    \includegraphics[width=0.3\textwidth]{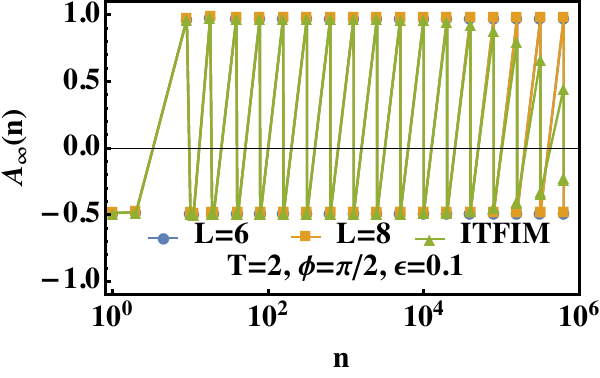}
    \includegraphics[width=0.29\textwidth]{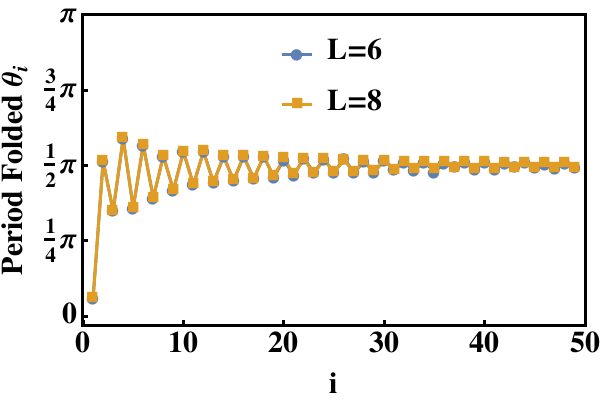}
    \includegraphics[width=0.19\textwidth]{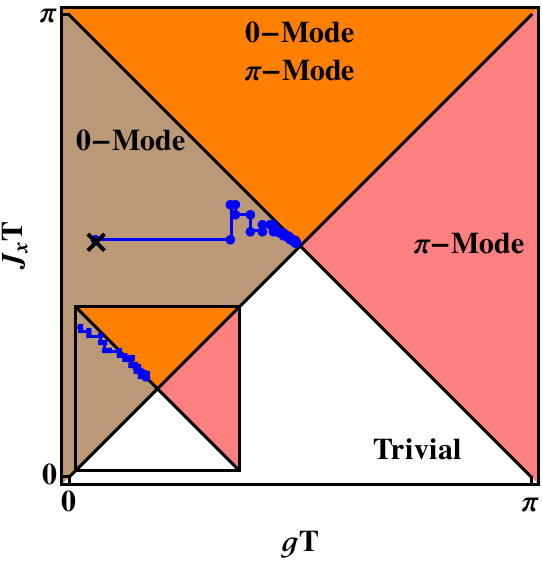}
    \includegraphics[width=0.3\textwidth]{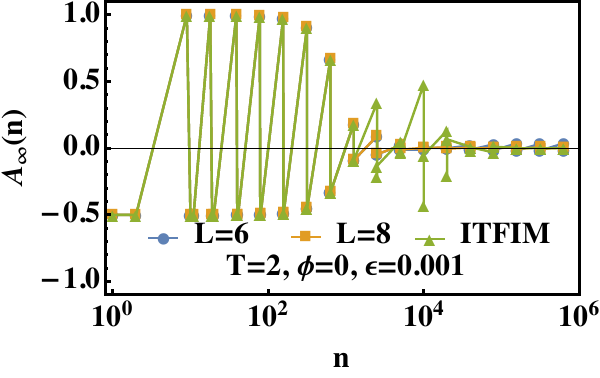}
    \includegraphics[width=0.29\textwidth]{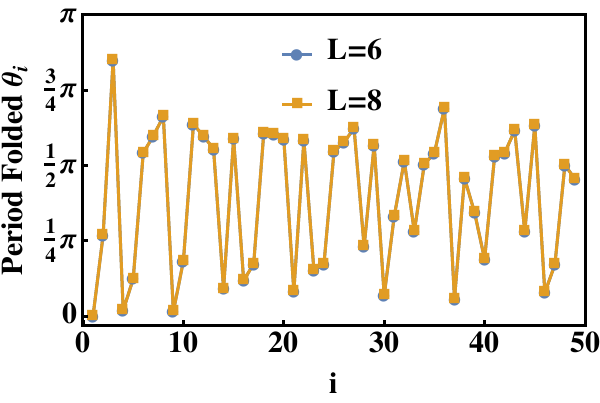}
    \includegraphics[width=0.19\textwidth]{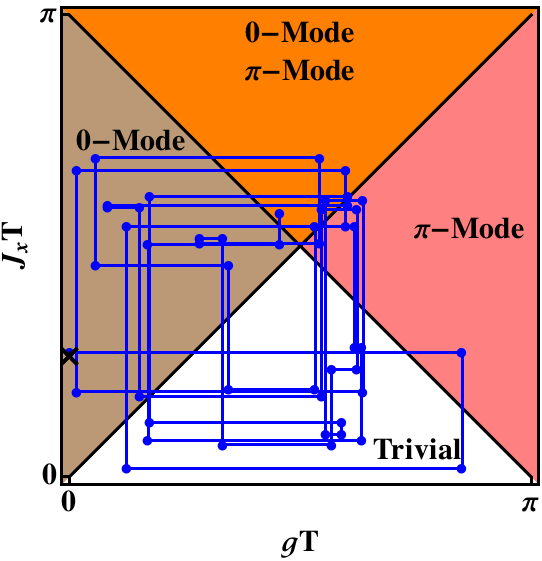}
    \caption{Infinite temperature autocorrelation functions from ED and from the ITFIM (left panels) for the $Z_3$ clock model \eqref{Eq: Z3 clock model} with $O_0 = (\sigma_1 + \sigma_1^\dagger)/\sqrt{2}$, $T=2$, $J=e^{i\phi}$ and $h = \epsilon + i2\pi/(3\sqrt{3})$. Also shown are  period folded Krylov angles constructed from $(U_2)^3$ (middle panels) and their trajectories in the phase diagram of the TFIM (see Appendix \ref{Sec:E} for the unfolded Krylov angles and the autocorrelaton functions on a linear scale). On period folding, the period-tripled mode behaves as a zero mode. The black cross (right panels) indicates the period folded Krylov angle on the first site. Top panels: the autocorrelation shows the presence of a long-lived edge mode for $\phi = \pi/2$,$\epsilon = 0.1$. Although the period folded Krylov angles approach $\pi/2$, they never fall outside the $0$-mode phase and the system still supports a long-lived period tripled edge mode. Bottom panels: the autocorrelation with tripled period decays much faster, and shows no long lived edge mode for $\phi=0, \epsilon = 0.001$. The period folded Krylov angles do not approach $\pi/2$ and the trajectory in the phase diagram initially starts from the $0$-mode phase and meanders through the phase diagram. $K_{\text{ITFIM}}$ is truncated to a $50 \times 50$ matrix and constructed from Krylov angles for $L=8$ with $\theta_{50}=\pi/2$. The ITFIM fails to capture the late-time dynamics in both cases in the left panels, indicating that more Krylov angles are needed to capture the late time behavior, see discussion in Appendix \ref{Sec:F}.}
    \label{Fig: Z3 clock model}
\end{figure*}

One can represent the mapping between the original model and the ITFIM in the phase diagram of the Floquet TFIM with homogeneous couplings, as shown in Fig.~\ref{Fig: Phase diagram}. After mapping to ITFIM, the Krylov angles $\{ \theta_i \}$ are identified as local transverse-fields, $g_i T = \theta_{2i-1}$, and local Ising couplings, $J_{xi}T = \theta_{2i}$. One constructs a series of data points, $(g_1T, J_{x1}T), (g_2T, J_{x1}T), (g_2T, J_{x2}T), (g_3T, J_{x2}T)\ldots$, and this trajectory is plotted in the phase diagram. In particular, one finds that the couplings from the edge to the bulk gradually deviate from the original phase in the unperturbed limit (triangles in Fig.~\ref{Fig: Phase diagram}). Eventually, this trajectory meanders in different phases and approaches $(\pi/2,\pi/2)$. This is consistent with Ref.~\cite{suchsland2023krylov} that $a_i, c_i \rightarrow 0$ and $b_i \rightarrow 1$ for large enough $i$ in chaotic systems, and hence $\theta_i \rightarrow \pi/2$ from \eqref{Eq: Krylov abc}. This also explains the long-lived quasi-stable mode observed in Fig.~\ref{Fig: perturbed TFIM}. In the language of ITFIM, there is a finite segment starting from the edge where the local couplings support the existence of the edge mode. After the first transient, the edge Majorana first decays to an approximate edge mode on this finite segment and survives for some time, but eventually decays into the bulk. Our results indicate that one need only keep the Krylov angles that contribute to the finite segment supporting edge modes in the ITFIM. For example, for the $0$-mode (blue circles in Fig.~\ref{Fig: Phase diagram}), we keep data points from the edge to the bulk site just before the first data point outside the $0$-mode phase, and construct the $\Tilde{K}_{\text{ITFIM}}$ by setting the last Krylov angle to be $\pi/2$. Note that $\Tilde{K}_{\text{ITFIM}}$ is a smaller matrix than $K_{\text{ITFIM}}$. The approximation of setting the final angle to $\pi/2$ both in $K_{\rm ITFIM}$ and $\tilde{K}_{\rm ITFIM}$ implies that once the particle reaches the last site it is lost to the bulk with unit probability, and there are no recurrences.
Thus, this approximation is equivalent to simulating a semi-infinite system.
We propose the following approximate autocorrelation function 
\begin{subequations}\label{Eq: Approx}
\begin{align}    &\Tilde{A}_\infty^\alpha(n) = |\Tilde{\psi}_{\alpha,1}|^2 e^{-\Gamma_\alpha n },\,\,
 \Tilde{A}_\infty^{0-\pi}(n) = \sum_{\alpha}\Tilde{A}_\infty^{\alpha}(n); \label{Eq: Approx A}\\ 
    &\Gamma_\alpha = \lim_{m \rightarrow \infty} \frac{-\ln[|(\Tilde{\psi}_{\alpha}|\Tilde{K}_{\text{ITFIM}}^m|\Tilde{\psi}_{\alpha})|]}{m},
    \label{Eq: Approx A-2}
\end{align}
\end{subequations}
where $\alpha = 0,\pi$ corresponds to $0$ or $\pi$-mode. $|\Tilde{\psi}_\alpha)$ is the approximate $\alpha$-mode  represented as a column vector in the Krylov basis and $\Tilde{\psi}_{\alpha,1}$ is it's first element. The approximate edge modes  can be analytically constructed as they are related to the Krylov angles as follows, see detailed derivation in Appendix \ref{Sec:C}.
\begin{align}
    &\frac{\Tilde{\psi}_{0,2i}}{\Tilde{\psi}_{0,2i-1}} = \tan\frac{\theta_{2i-1}}{2}; &\frac{\Tilde{\psi}_{0,2i+1}}{\Tilde{\psi}_{0,2i}} = \cot\frac{\theta_{2i}}{2};\nonumber\\
    &\frac{\Tilde{\psi}_{\pi,i+1}}{\Tilde{\psi}_{\pi,i}} = -\cot\frac{\theta_i}{2}.
\end{align}
The decay rate $\Gamma_\alpha$ defined in \eqref{Eq: Approx A-2} is based on the approximation, $\exp(-\Gamma_\alpha m) \approx |(\Tilde{\psi}_{\alpha}|\Tilde{K}_{\text{ITFIM}}^m|\Tilde{\psi}_{\alpha})|$, and we set $m=1000$ in the numerical computation. For the $0$-$\pi$ phase where both modes are present, the upper envelope of the autocorrelation can be treated as the sum of two approximate autocorrelations, c.f. \eqref{Eq: Approx A}.
The numerical results of the approximate autocorrelations are plotted in the top panels of Fig.~\ref{Fig: perturbed TFIM} and give good qualitative agreement with ED.

In the last example, $Z_3$ clock model \eqref{Eq: Z3 clock model}, we consider $O_0 = (\sigma_1 + \sigma_1^\dagger)/\sqrt{2}$ with $T=2$, $J=e^{i\phi}$, $h = \epsilon + i2\pi/(3\sqrt{3})$. The $Z_3$ clock model has an edge mode completely localized on the first site for $\epsilon=0$, where it shows period tripled dynamics, and is closely related to the decoupled edge Majorana in the TFIM for $g=0$ \cite{sreejith2016parafermion,Fendley2012}. As one moves away from $\epsilon=0$, the existence of the edge mode depends on the phase $\phi$ of $J$   \cite{sreejith2016parafermion,Fendley2012,jermyn2014stability}. In Fig.~\ref{Fig: Z3 clock model}, we present two setups: (i) long-lived, period tripled edge mode  at $\phi=\pi/2$, $\epsilon =0.1$ (top panels). (ii) 
short-lived, period tripled edge mode at $\phi=0$, $\epsilon = 0.001$. We numerically compute the autocorrelation at three consecutive times, $n,n+1, n+2$, where $n$ equally slices the time in the logarithmic-scale and $n = 1\ \text{mod}\ 3$. The upper (lower) envelope corresponds to $n+2\,  (n,n+1)$, the linear-scale plots are presented in Appendix \ref{Sec:F}. For a better understanding of the period-tripled mode in the phase diagram of the TFIM, we also consider the period folded Krylov angles obtained from $(U_2)^3$. This is equivalent to observing the dynamics every three periods, smearing out the period-tripled oscillations such that the edge mode behaves effectively as a $0$-mode. The  period folded Krylov angles show qualitatively different behavior from the first model $U_{1}$. In setup (i), the Krylov angles approach $\pi/2$ in the bulk, even though the autocorrelation has not decayed. This is different from the chaotic cases in Fig.~\ref{Fig: Phase diagram} because for (i) the period folded Krylov angles are all still within the $0$-mode phase, thus supporting the long-lived mode. In contrast to (i),  in setup (ii), the trajectory of the period folded Krylov angles starts from the $0$-mode phase but does not approach the center of the phase diagram. Hence, the approximate autocorrelation function \eqref{Eq: Approx A} cannot be used here, see discussion in Appendix \ref{Sec:F}. For both cases  $K_{\rm ITFIM}$ truncated to $N=50$, fails to describe the autocorrelation at late-times indicating that the clock model requires more Krylov angles to faithfully capture the late time dynamics, see plots in Appendix \ref{Sec:F}.

\section{Conclusion} \label{Sec5}
We showed that the stroboscopic dynamics of any Hermitian operator under any Floquet unitary in any spatial dimension in Krylov space, is identical to the dynamics of the edge operator of a Floquet ITFIM in one spatial dimension. The details of the operator and the Floquet unitary determine the precise values of the inhomogeneous couplings of the ITFIM. The latter is a non-interacting model that can be written in terms of Majorana bilinears. Thus, this mapping allows one to leverage methods from single-particle quantum mechanics to the study of operator dynamics of an interacting quantum problem. We presented applications involving interacting Floquet spin $1/2$
chains (see also Appendix \ref{Sec:D}) and the $Z_3$ clock model, where the latter hosts period-tripled edge modes. The operator dynamics  shows certain universal features captured by how the Krylov parameters vary in the topological phase diagram of the non-interacting model. This universality allowed us to capture the dynamics by an efficient truncation in Krylov space. Future directions would involve further exploration of the $Z_3$ clock model and models in  spatial dimensions $d>1$, as well as the study of other metrics such as operator entanglement and out of time ordered correlators. 

\emph{Acknowledgments}: This work was supported by the US Department of Energy, Office of
Science, Basic Energy Sciences, under Award No.~DE-SC0010821. HCY acknowledges support of the NYU IT High Performance Computing resources, services, and staff expertise. 

\appendix
\begin{widetext}
\section{Operator Krylov space Floquet unitary}
\label{Sec:A}
Here we provide the detailed derivation of the relation between the matrix elements of the operator Krylov space Floquet unitary generated by the Arnoldi iteration. Starting with an initial normalized Hermitian operator $O_0$, one can generate a set of operators via the Floquet unitary $U$: $\{|O_0), |O_1), |O_2), \ldots \}$, where $|O_i)$ is the vector representation of $O_i$ with $|O_i) = K^i|O_0) = (U^\dagger)^i O_0 U^i$. Since the seed operator $O_0$ is Hermitian, and due to time-evolution by a Floquet unitary, all  $\{O_i \}$ are Hermitian as well. The inner product of two operators is defined as $(A|B) = \text{Tr}[A^\dagger B]/N$ with $A$ and $B$ being $N \times N$ matrices. From this definition of the inner product, one can show that $(O_i|K|O_j) = (O_{i-1}|O_j)$ by using cyclic property of the trace and $UO_iU^\dagger = O_{i-1}$. Therefore, one obtains the relation: $(O_i|K = (O_{i-1}|$. One can perform the Gram-Schmidt procedure to obtain the orthonormal basis $\{|\mathcal{O}_0), |\mathcal{O}_1), |\mathcal{O}_2), \ldots \}$. The goal is to find a representation of $K$ in Krylov space (in the Gram-Schmidt orthonormal basis). According to \cite{suchsland2023krylov}, $K$ can be parameterized by three series of numbers $\{a_i\}, \{b_i\}$ and $\{c_i \}$ as follows
\begin{align}
&b_i=(\mathcal{O}_i|K|\mathcal{O}_{i-1});
&\frac{a_i}{c_i}c_j=(\mathcal{O}_i|K|\mathcal{O}_j)\ \text{for}\ i \leq j,
\end{align}
where $a_0 = c_0$ and $(\mathcal{O}_i|K|\mathcal{O}_j)=0$ for $i > j+1$. For example, $K$ has the following explicit expression for $N = 6$
\begin{align}
    \begin{pmatrix}
        a_0 & c_1 & c_2 & c_3 & c_4 & c_5\\
        b_1 & a_1 & \frac{a_1}{c_1}c_2 & \frac{a_1}{c_1}c_3 & \frac{a_1}{c_1}c_4 & \frac{a_1}{c_1}c_5\\
        0 & b_2 & a_2 & \frac{a_2}{c_2}c_3 & \frac{a_2}{c_2}c_4 & \frac{a_2}{c_2}c_5\\
        0 & 0 & b_3 & a_3 & \frac{a_3}{c_3}c_4 & \frac{a_3}{c_3}c_5\\
        0 & 0 & 0 & b_4 & a_4 & \frac{a_4}{c_4}c_5\\
        0 & 0 & 0 & 0 & b_5 & a_5
    \end{pmatrix}.
\end{align}
Here we follow the proof in Ref.~\cite{suchsland2023krylov}. Since the the Gram-Schmidt procedure constructs an orthogonal basis iteratively, $|\mathcal{O}_i)$ is a linear combination of $\{ |O_t) \}$ up to $t = i$,
\begin{align}
    |\mathcal{O}_i) = \sum_{t=0}^i \alpha_{i,t}|O_t).
\end{align}
Explicitly, one obtains a set of equations
\begin{align}
    |\mathcal{O}_0) &=\alpha_{0,0} |O_0)\nonumber\\
    |\mathcal{O}_1) &=\alpha_{1,0} |O_0) + \alpha_{1,1} |O_1)\nonumber\\
    |\mathcal{O}_2) &=\alpha_{2,0} |O_0) + \alpha_{2,1} |O_1)+ \alpha_{2,2} |O_2)\nonumber\\
    &\ \vdots
\end{align}
The inverse of the above relations are
\begin{align}
    |O_0) &=\frac{1}{\alpha_{0,0}} |\mathcal{O}_0)\nonumber\\
    |O_1) &=-\frac{\alpha_{1,0}}{\alpha_{1,1}} |O_0) + \frac{1}{\alpha_{1,1}} |\mathcal{O}_1) = -\frac{\alpha_{1,0}}{\alpha_{0,0}\alpha_{1,1}} |\mathcal{O}_0)+\frac{1}{\alpha_{1,1}} |\mathcal{O}_1)\nonumber\\
    |O_2) &= -\frac{\alpha_{2,0}}{\alpha_{2,2}}|O_0) -\frac{\alpha_{2,1}}{\alpha_{2,2}}|O_1) + \frac{1}{\alpha_{2,2}}|\mathcal{O}_2) = \frac{\alpha_{2,1}\alpha_{1,0}-\alpha_{2,0}\alpha_{1,1}}{\alpha_{0,0}\alpha_{1,1}\alpha_{2,2}}|\mathcal{O}_0) -\frac{\alpha_{2,1}}{\alpha_{1,1}\alpha_{2,2}}|\mathcal{O}_1) + \frac{1}{\alpha_{2,2}}|\mathcal{O}_2)\nonumber\\
    &\ \vdots
\end{align}
The above is equivalent to the inverse relation
\begin{align}
    |O_t) = \sum_{i=0}^t (\alpha^{-1})_{t,i}|\mathcal{O}_i),\label{eq:gs1}
\end{align}
where $\alpha$ and $\alpha^{-1}$ are lower triangular matrices. From this relation, one concludes $(\mathcal{O}_i|O_t) = 0$ for $i > t$. Now, we examine the matrix element $(\mathcal{O}_i|K|\mathcal{O}_j)$ for different $i,j$. First, for $i > j+1$,
\begin{align}
    (\mathcal{O}_i|K|\mathcal{O}_j) = \sum_{t=0}^j \alpha_{j,t}(\mathcal{O}_i|K|O_t) =  \sum_{t=0}^j \alpha_{j,t}(\mathcal{O}_i|O_{t+1}) = 0,
\end{align}
where we apply the fact that $(\mathcal{O}_i|O_{t+1}) = 0$ since $i > j+1 \geq t+1$. Second, for $i \leq j$,
\begin{align}
    (\mathcal{O}_i|K|\mathcal{O}_j) = \sum_{t=0}^i \alpha^*_{i,t}(O_t|K|\mathcal{O}_j) =  \alpha^*_{i,0} (O_0|K|\mathcal{O}_j) + \sum_{t=1}^i \alpha^*_{i,t}(O_{t-1}|\mathcal{O}_j) = \alpha^*_{i,0}c_j,
\end{align}
where we apply $c_j = (\mathcal{O}_0|K|\mathcal{O}_j) = (O_0|K|\mathcal{O}_j)$, and $(O_{t-1}|\mathcal{O}_j) = 0$ for $j > i-1 \geq t-1$. The latter follows from \eqref{eq:gs1} and the fact that $|\mathcal{O}_j)$ are Gram-Schmidt orthogonalized. To determine $\alpha^*_{i,0}$, one considers the $i = j$ case,
\begin{align}
    a_i = (\mathcal{O}_i|K|\mathcal{O}_i) = \alpha^*_{i,0}c_i.
\end{align}
One concludes $\alpha^*_{i,0} = a_i/c_i$ and derive
\begin{align}
    (\mathcal{O}_i|K|\mathcal{O}_j) = \frac{a_i}{c_i}c_j.
\end{align}
For $i = j+1$, there are no further constraints, and we simply define it as $b_i = (\mathcal{O}_i|K|\mathcal{O}_{i-1})$. Therefore, we prove the expression of $K$ above. We emphasize that the above form does require unitary evolution because we used the condition: $K|O_t) = |O_{t+1})$ and $(O_t|K = (O_{t-1}|$.

If the initial operator $O_0$ is Hermitian, $\{|\mathcal{O}_0), |\mathcal{O}_1), |\mathcal{O}_2), \ldots \}$ are all Hermitian, which implies that $\{a_i\}, \{b_i\}$ and $\{c_i \}$ are all real numbers. This can be shown directly from the cyclic property of the trace
\begin{align}
    (\mathcal{O}_i|K|\mathcal{O}_j)^* = \frac{1}{N}\text{Tr}[(\mathcal{O}_i U^\dagger \mathcal{O}_j U)^\dagger] = \frac{1}{N}\text{Tr}[ U^\dagger \mathcal{O}_j U\mathcal{O}_i] = \frac{1}{N}\text{Tr}[\mathcal{O}_i U^\dagger \mathcal{O}_j U] = (\mathcal{O}_i|K|\mathcal{O}_j).
\end{align}
Since $(\mathcal{O}_i|K|\mathcal{O}_j)$ is real, $\{a_i\}, \{b_i\}$ and $\{c_i \}$ are real as well. Moreover,
the requirement of unitarity, $K^\intercal K = \mathbb{I}$, gives further constraints on $\{a_i\}, \{b_i\}$ and $\{c_i \}$. For example, for $N = 4$,
\begin{align}
K^\intercal K = 
    \begin{pmatrix}
  a_0^2+b_1^2 & c_1 \left(\frac{a_1 b_1}{c_1}+a_0\right) & c_2 \left(\frac{a_1 b_1}{c_1}+a_0\right) & c_3 \left(\frac{a_1
   b_1}{c_1}+a_0\right) \\
 c_1 \left(\frac{a_1 b_1}{c_1}+a_0\right) & a_1^2+b_2^2+c_1^2 & c_2
   \left(\frac{a_2 b_2}{c_2}+c_1\left(1+\frac{a_1^2}{c_1^2}\right)\right) & c_3
   \left(\frac{a_2 b_2}{c_2}+c_1\left(1+\frac{a_1^2}{c_1^2}\right)\right) \\
 c_2 \left(\frac{a_1 b_1}{c_1}+a_0\right) & c_2
   \left(\frac{a_2 b_2}{c_2}+c_1\left(1+\frac{a_1^2}{c_1^2}\right)\right) & a_2^2+b_3^2+
   \left(1+\frac{a_1^2}{c_1^2}\right)c_2^2 & c_3\left(
   \frac{a_3 b_3}{c_3}+c_2 \left(1+ \frac{a_1^2}{c_1^2} + \frac{a_2^2}{c_2^2}\right)\right) \\
 c_3 \left(\frac{a_1 b_1}{c_1}+a_0\right) & c_3
   \left(\frac{a_2 b_2}{c_2}+c_1\left(1+\frac{a_1^2}{c_1^2}\right)\right) & c_3\left(
   \frac{a_3 b_3}{c_3}+c_2 \left(1+ \frac{a_1^2}{c_1^2} + \frac{a_2^2}{c_2^2}\right)\right) &a_3^2 + 
   \left(1+\frac{a_1^2}{c_1^2}+\frac{a_2^2}{c_2^2}\right)c_3^2 \\
    \end{pmatrix}.
\end{align}
The general structure of $K^\intercal K$ represented as a $N \times N$ matrix is summarized as follows (using $a_0=c_0$):
\begin{align}
    &(K^\intercal K)_{1,1} = a_0^2 + b_1^2;\\
    &(K^\intercal K)_{i,i} = a_{i-1}^2 + b_i^2 + c_{i-1}^2\sum_{l = 0}^{i-2}\frac{a_l^2}{c_l^2},\ \text{for}\ 2\leq i \leq N-1;\\
    &(K^\intercal K)_{N,N} = a_{N-1}^2 + c_{N-1}^2\sum_{l = 0}^{N-2}\frac{a_l^2}{c_l^2};\\
    &(K^\intercal K)_{i,j} = c_{j-1} \left( \frac{a_i b_i}{c_i} + c_{i-1} \sum_{l = 0}^{i-1}\frac{a_l^2}{c_l^2}\right),\ \text{for}\ i < j\ \text{(upper triangle)};\\
    &(K^\intercal K)_{i,j} = c_{i-1} \left( \frac{a_j b_j}{c_j} + c_{j-1} \sum_{l = 0}^{j-1}\frac{a_l^2}{c_l^2}\right),\ \text{for}\ i > j\ \text{(lower triangle)}.
\end{align}
By setting $K^\intercal K = \mathbb{I}$, one can solve for $\{a_i\}$ and $\{c_i\}$ in terms of $\{b_i\}$ order by order in $i$. A computation performed to $i$-th order for example 
requires knowledge of the first $i$ coefficients $b_i$.
Since $K$ preserves unitarity, $b_i = (\mathcal{O}_i|K|\mathcal{O}_{i-1})$ can be considered as an inner product of two normalized operators, $|\mathcal{O}_i)$ and $K|\mathcal{O}_{i-1})$. Therefore, $|b_i| \leq 1$. Similarly, $|a_i| \leq 1$ and $|c_i| \leq 1$ as these are also  inner products between two normalized operators. 
It is convenient to parameterize 
\begin{align}
b_i = \sin\theta_i,
\end{align}
in the following derivation. Note that $\theta_i \in [0, \pi]$, since one can show that $b_i$ is non-negative 
\begin{align}
    b_i=(\mathcal{O}_i|K|\mathcal{O}_{i-1}) = \sqrt{(\mathcal{O}_i^\prime|\mathcal{O}_i^\prime)} \geq 0,
\end{align}
where $\mathcal{O}_i'$ is constructed from the Gram-Schmidt procedure
\begin{align}
    &|\mathcal{O}_i^\prime) = K|\mathcal{O}_{i-1}) - \sum_{j=1}^{i-1}|\mathcal{O}_j)(\mathcal{O}_j|K|\mathcal{O}_{i-1});
    &|\mathcal{O}_i) = \frac{|\mathcal{O}_i^\prime)}{\sqrt{(\mathcal{O}_i^\prime|\mathcal{O}_i^\prime)}}.
\end{align}

In 1st order,  using $(K^\intercal K)_{1,1} = 1$ gives 
\begin{align}
    a_0 = c_0 = \pm \sqrt{1-b_1^2} =  \pm |\cos\theta_1| = \cos\theta_1,
\end{align}
where we have  absorbed the $\pm$ into the definition of $\theta_1$. One can always find a $\theta_1$ such that it leaves $b_1$ unchanged but affects the sign of $a_0$ and $c_0$ since $\cos{(\pi - \theta_1)} = -\cos{\theta_1}$ and $b_1 = \sin(\theta_1) = \sin(\pi-\theta_1)$.
In 2nd order, using $(K^\intercal K)_{2,2} = 1$ and $(K^\intercal K)_{1,i>1} = (K^\intercal K)_{i>1,1} = 0$, gives
\begin{align}
    &a_1^2 +c_1^2 = 1-b_2^2 = \cos^2\theta_2;&\frac{a_1}{c_1} = -\frac{c_0}{b_1} =  -\cot\theta_1.
\end{align}
Therefore, one obtains
\begin{align}
    &a_1 =  \cos\theta_1\cos\theta_2; &c_1 =  -\sin\theta_1\cos\theta_2.
\end{align}
In 3rd order, using $(K^\intercal K)_{3,3} = 1$ and $(K^\intercal K)_{2,i>2} = (K^\intercal K)_{i>2,2} = 0$, gives
\begin{align}
    &a_2^2 + \left( 1 + \frac{a_1^2}{c_1^2}\right) c_2^2 = 1-b_3^2; &\frac{a_2}{c_2} = -\frac{c_1}{b_2}\left( 1 + \frac{a_1^2}{c_1^2}\right).
\end{align}
Using $a_1^2+c_1^2 = 1-b_2^2 = \cos^2\theta_2$ and $b_3 = \sin\theta_3$, one obtains
\begin{align}
    &a_2^2 + \cos^2\theta_2 \left(\frac{c_2}{c_1}\right)^2  = \cos^2\theta_3; &\frac{a_2}{c_2/c_1} = -\frac{\cos\theta_2}{\tan\theta_2}.
\end{align}
The solutions are
\begin{align}
    &a_2 =\cos\theta_2\cos\theta_3; &\frac{c_2}{c_1}  = -\tan\theta_2\cos\theta_3.
\end{align}
In 4th order, using $(K^\intercal K)_{4,4} = 1$ and $(K^\intercal K)_{3,i>3} = (K^\intercal K)_{i>3,3} = 0$, gives 
\begin{align}
    &a_3^2 + \left( 1 + \frac{a_1^2}{c_1^2}+\frac{a_2^2}{c_2^2}\right) c_3^2 = 1-b_4^2; &\frac{a_3}{c_3} = -\frac{c_2}{b_3}\left( 1 + \frac{a_1^2}{c_1^2} + \frac{a_2^2}{c_2^2}\right).
\end{align}
Using $1 + (a_1^2/c_1^2)+(a_2^2/c_2^2) = \cos^2\theta_3/c_2^2$ and $b_4 = \sin\theta_4$, we simplify the equations as
\begin{align}
    &a_3^2 +  \cos^2\theta_3 \left(\frac{c_3}{c_2}\right)^2  = \cos^2\theta_4; &\frac{a_3}{c_3/c_2} =  -\frac{\cos\theta_3}{\tan\theta_3}.
\end{align}
The solutions are
\begin{align}
    &a_3  = \cos\theta_3\cos\theta_4; &\frac{c_3}{c_2} = -\tan\theta_3\cos\theta_4.
\end{align}
The same pattern appears in higher order as well. At $(i+1)$-th order with $i+1 < N$, the solutions are
\begin{align}
    &a_i = \cos\theta_i\cos\theta_{i+1}; &\frac{c_i}{c_{i-1}} = -\tan\theta_{i}\cos\theta_{i+1}.
\end{align}
At $N$-th order, the solutions still obey the recursive relation by setting $b_{N} = 0\ (\theta_N = 0)$
\begin{align}
    &a_{N-1} = \cos\theta_{N-1}; & \frac{c_{N-1}}{c_{N-2}} = -\tan\theta_{N-1}.
\end{align}
In summary, we obtain the following analytic expressions for $\{a_i\}, \{b_i\}$ and $\{c_i \}$,
\begin{align}
    &a_i = \cos\theta_i\cos\theta_{i+1};\quad\quad\quad\quad\quad\quad
    b_i = \sin\theta_i;
    &c_i = (-1)^i\cos\theta_{i+1} \prod_{j=1}^i\sin\theta_j,
    \label{Eq; Krylov theta}
\end{align}
with $\theta_0 = \theta_{N} = 0$.

\section{Mapping to a Floquet Inhomogeneous transverse-field Ising model (ITFIM)}
\label{Sec:B}
In this section, we present a universal model generating the same Krylov coefficients \eqref{Eq; Krylov theta}. In particular, we show that
identical dynamics is generated by the Floquet transverse-field Ising model with inhomogenous couplings  and open boundary conditions, which we dub the "inhomogenous transverse field Ising model" (ITFIM),
\begin{align}
&U_{\text{ITFIM}} = U_z U_{xx};\quad\quad\quad\quad\quad\quad
U_z = \prod_{j=1}^{N/2} \exp\left(-i\frac{\theta_{2j-1}}{2}\sigma_j^z \right);
&U_{xx} = \prod_{j=1}^{(N-2)/2} \exp\left(-i\frac{\theta_{2j}}{2}\sigma_j^x \sigma_{j+1}^x \right).
\end{align}
Above, we assume $N$ is even and $\theta_i \in [0,\pi]$.  
The above unitary is bilinear in terms of Majorana fermions defined as
\begin{align}
    &\gamma_{2l-1} = \sigma_l^x\prod_{j=1}^{l-1}\sigma_j^z ; & \gamma_{2l} = \sigma_l^y\prod_{j=1}^{l-1}\sigma_j^z .
\end{align}
In particular, $-i\gamma_{2l-1}\gamma_{2l} = \sigma^z_l, -i\gamma_{2l}\gamma_{2l+1} =\sigma^x_l\sigma^x_{l+1} $. One can show that the Majoranas evolve under unitary evolution as follows: Under $U_z$,
\begin{align}
    &U_z^\dagger \gamma_{2l-1} U_z = \cos(\theta_{2l-1})\gamma_{2l-1} - \sin(\theta_{2l-1})\gamma_{2l};
    &U_z^\dagger \gamma_{2l} U_z = \sin(\theta_{2l-1})\gamma_{2l-1} + \cos(\theta_{2l-1})\gamma_{2l}.
\end{align}
Under $U_{xx}$,
\begin{align}
    &U_{xx}^\dagger \gamma_1 U_{xx} = \gamma_1; &U_{xx}^\dagger \gamma_N U_{xx} = \gamma_N;\\
    &U_{xx}^\dagger \gamma_{2l} U_{xx} = \cos(\theta_{2l})\gamma_{2l} -\sin(\theta_{2l})\gamma_{2l+1}; &U_{xx}^\dagger \gamma_{2l+1} U_{xx} = \sin(\theta_{2l})\gamma_{2l} + \cos(\theta_{2l})\gamma_{2l+1}. 
\end{align}

Let us consider an operator $\Psi$ that can be written as a linear combination of Majoranas, $\Psi = \sum_{i} \psi_i\gamma_i$.  The coefficients can be viewed as a column vector $\vec{\psi} = (\psi_1, \psi_2, \psi_3, \ldots)^\intercal$. Under the effect of the two unitaries, the coefficients change according to, $U_z^\dagger \Psi U_z = \sum_{k} \psi_k' \gamma_k$ with $\vec{\psi}' = \overline{K}_z \vec{\psi}$  and $U_{xx}^\dagger \Psi U_{xx} = \sum_{k} \psi_k'' \gamma_k$ with $\vec{\psi}'' = \overline{K}_{xx} \vec{\psi}$. For example for $N=6$
\begin{align}
&\overline{K}_z = \begin{pmatrix}
         \cos\theta_1 & \sin\theta_1 & 0 & 0 & 0 & 0 \\
 -\sin\theta_1 & \cos \theta_1 & 0 & 0 & 0 & 0 \\
 0 & 0 & \cos \theta_3 & \sin \theta_3 & 0 & 0 \\
 0 & 0 & -\sin \theta_3 & \cos \theta_3 & 0 & 0 \\
 0 & 0 & 0 & 0 & \cos \theta_5 & \sin \theta_5 \\
 0 & 0 & 0 & 0 & -\sin \theta_5 & \cos \theta_5
\end{pmatrix};\\
&\overline{K}_{xx}=
\begin{pmatrix}
    1 & 0 & 0 & 0 & 0 & 0 \\
 0 & \cos \theta_2 & \sin \theta_2 & 0 & 0 & 0 \\
 0 & -\sin \theta_2 & \cos \theta_2 & 0 & 0 & 0 \\
 0 & 0 & 0 & \cos \theta_4 & \sin \theta_4 & 0 \\
 0 & 0 & 0 & -\sin \theta_4 & \cos \theta_4 & 0 \\
 0 & 0 & 0 & 0 & 0 & 1
\end{pmatrix}.
\end{align}
The full unitary evolution in the Majorana basis is
\begin{align}
    \overline{K}_{\text{ITFIM}} &= \overline{K}_{xx} \overline{K}_z\\
    &= \begin{pmatrix}
        \cos \theta_1 & \sin \theta_1 & 0 & 0 & 0 & 0 \\
 -\sin \theta_1 \cos \theta_2 & \cos
   \theta_1 \cos \theta_2 & \sin \theta_2
   \cos \theta_3 & \sin \theta_2 \sin \theta_3 & 0 & 0 \\
 \sin \theta_1 \sin \theta_2 & -\cos \theta_1\sin \theta_2 & \cos \theta_2 \cos
   \theta_3 & \cos \theta_2\sin \theta_3  & 0
   & 0 \\
 0 & 0 & -\sin \theta_3 \cos \theta_4 & \cos
   \theta_3 \cos \theta_4 & \sin \theta_4
   \cos \theta_5 & \sin \theta_4 \sin \theta_5 \\
 0 & 0 & \sin \theta_3 \sin \theta_4 & -\cos \theta_3\sin \theta_4  & \cos \theta_4
   \cos \theta_5 & \cos \theta_4 \sin \theta_5  \\
 0 & 0 & 0 & 0 & -\sin \theta_5 & \cos \theta_5
\end{pmatrix}.
\end{align}
One can also represent $\overline{K}_{\text{ITFIM}}$ in Krylov space where the Gram-Schmidt orthonormal basis is generated from $|O_0) = |\gamma_1) = (1,0,0,\ldots)^\intercal$. We explicitly show the first few steps in this procedure. First,  $|O_0) = (1,0,0,\ldots)^\intercal$, and one can directly compute 
$a_0 = (O_0|\overline{K}_{\text{ITFIM}}|O_0) = \cos\theta_1$. Next, one constructs the new operator,
$|O_1) = \overline{K}_{\text{ITFIM}}|O_0)$,
\begin{align}
    |O_1) = (\cos\theta_1, -\sin\theta_1\cos\theta_2, \sin\theta_1\sin\theta_2,0,\ldots)^\intercal.
\end{align}
By Gram-Schmidt, one obtains the new orthonormal state $|\mathcal{O}_1)$,
\begin{align}
    |\mathcal{O}_1) = (0,-\cos\theta_2,\sin\theta_2,0,\ldots)^\intercal.
\end{align}
Now, one can complete the upper $2\times 2$ block of $K_{\text{ITFIM}}$,
\begin{align}
    &a_1 = (\mathcal{O}_1|\overline{K}_{\text{ITFIM}}|\mathcal{O}_1)  = \cos\theta_1\cos\theta_2;\quad
    b_1 = (\mathcal{O}_1|\overline{K}_{\text{ITFIM}}|O_0) = \sin\theta_1;
    &c_1 = (O_0|\overline{K}_{\text{ITFIM}}|\mathcal{O}_1) = -\sin\theta_1\cos\theta_2.
\end{align}
For further steps, we use symbolic computation on Mathematica. 
In particular, one can show explicitly for $N=6$,
\begin{align}
    K_{\text{ITFIM}} = \begin{pmatrix}
        a_0 & c_1 & c_2 & c_3 & c_4 & c_5\\
        b_1 & a_1 & \frac{a_1}{c_1}c_2 & \frac{a_1}{c_1}c_3 & \frac{a_1}{c_1}c_4 & \frac{a_1}{c_1}c_5\\
        0 & b_2 & a_2 & \frac{a_2}{c_2}c_3 & \frac{a_2}{c_2}c_4 & \frac{a_2}{c_2}c_5\\
        0 & 0 & b_3 & a_3 & \frac{a_3}{c_3}c_4 & \frac{a_3}{c_3}c_5\\
        0 & 0 & 0 & b_4 & a_4 & \frac{a_4}{c_4}c_5\\
        0 & 0 & 0 & 0 & b_5 & a_5
    \end{pmatrix},
\end{align}
where
\begin{align}
    &a_i = \cos\theta_i\cos\theta_{i+1};\quad\quad\quad\quad\quad\quad
    b_i = \sin\theta_i;
    &c_i = (-1)^i\cos\theta_{i+1} \prod_{j=1}^i\sin\theta_j,
\end{align}
with $\theta_0 = \theta_{N} = 0$. This is exactly the same as the parameterization in \eqref{Eq; Krylov theta}. Thus, one can map the Floquet dynamics of any Hermitian operator in Krylov space to the dynamics of the edge Majorana $\gamma_1$ of an inhomogeneous Floquet transverse-field Ising model.

Numerically, one starts with some generic model and an initial Hermitian operator. One only needs to compute the  $\{a_i\}$ series in order to determine the corresponding parameterization angles $\{\theta_i\}\in \left[0,\pi\right]$, and this contains all the information in Krylov space.  

Note that $ \cos\theta_{2l-1} = (a_{2l-2}a_{2l-4}\ldots a_{0})/(a_{2l-3}a_{2l-5}\ldots a_1)$ and $\cos\theta_{2l} = (a_{2l-1}a_{2l-3}\ldots a_{1})/(a_{2l-2}a_{2l-4}\ldots a_0)$ with $\cos\theta_1 = a_0$. The Krylov angles can be solved iteratively and uniquely since all angles lie within $[0,\pi]$. 
However, it is impossible to numerically exhaust all $\{a_i\}$, which are typically exponentially many. Practically, one truncates $K$ into a $M \times M$ matrix. In chaotic system, $\theta_i$ approaches $\pi/2$ for large $i$. Therefore, one imposes the condition $\theta_M = \pi/2$ for the truncated $K$. This truncation will violate the unitarity of $K$. In particular, in the Krylov space orthonormal basis, $ K^\intercal K $ is still a diagonal matrix, but with matrix elements $(1, \ldots, 1, 0)$ along the diagonal. In practice, for sufficiently large $M$, physical observables, such as autocorrelation functions, will not be sensitive to the truncation. This truncation where the last Krylov angle is set to $\pi/2$ is
equivalent to approximating the bulk of the chain as an ideal reservoir because once the particle reaches the last site, it does not reflect back, but hops to the bulk with unit probability. Thus, despite the Krylov matrix being of finite size, setting the last angle to $\pi/2$ eliminates recurrences, and is equivalent to simulating a semi-infinite system.

\section{Approximate $0/\pi$ edge modes and their decay rate}
\label{Sec:C}
Here we provide the calculation of approximate edge modes for a given set of Krylov angles. In addition, we discuss the numerical computation of the decay rate. To derive the solution of the edge mode, it is easier to look for the left eigenvector of $K$
\begin{align}
    (\psi|K = \lambda(\psi|.
\end{align}
More explicitly,
\begin{align}
    \begin{pmatrix}
        \psi_1 & \psi_2 & \psi_3 & \psi_4 & \psi_5 &  \ldots
    \end{pmatrix}
    \begin{pmatrix}
        a_0 & c_1 & c_2 & c_3 & c_4 & \ldots\\
        b_1 & a_1 & \frac{a_1}{c_1}c_2 & \frac{a_1}{c_1}c_3 & \frac{a_1}{c_1}c_4 & \ldots\\
        0 & b_2 & a_2 & \frac{a_2}{c_2}c_3 & \frac{a_2}{c_2}c_4 & \ldots\\
        0 & 0 & b_3 & a_3 & \frac{a_3}{c_3}c_4 & \ldots\\
        0 & 0 & 0 & b_4 & a_4 & \ldots\\
        \vdots & \ddots & \ddots & \ddots & \ddots & \ddots
    \end{pmatrix}
    = \lambda\begin{pmatrix}
        \psi_1 & \psi_2 & \psi_3 & \psi_4 & \psi_5 & \ldots
    \end{pmatrix},
\end{align}
where the coefficients of $(\psi|$ can be solved column by column and $\lambda = 1$ ($\lambda = -1$) corresponds to a $0$-mode ($\pi$-mode). Let us start with $\lambda = 1$. From the first column of $K$, one obtains
\begin{align}
    \cos\theta_1\psi_1 + \sin\theta_1\psi_2  = \psi_1,
\end{align}
where we have used the Krylov angle parameterization of $\{ a_i \}$, $\{ b_i \}$ and $\{ c_i \}$. The ratio of the first two coefficients is
\begin{align}
    \frac{\psi_2}{\psi_1} = \frac{1-\cos\theta_1}{\sin\theta_1} = \tan\frac{\theta_1}{2}.
\end{align}
From the second column of $K$, one obtains 
\begin{align}
    -\sin\theta_1\cos\theta_2 \psi_1 + \cos\theta_1\cos\theta_2 \psi_2 + \sin\theta_2 \psi_3 = \psi_2.
\end{align}
With $\psi_2/\psi_1 = \tan(\theta_1/2)$, one obtains
\begin{align}
    \frac{\psi_3}{\psi_2} = \frac{1+\cos\theta_2}{\sin\theta_2} = \cot\frac{\theta_2}{2}.
\end{align}
Other columns also obey this alternating pattern and one obtains the analytic expression of the $0$-mode 
\begin{align}
    &\frac{\psi_{2i}}{\psi_{2i-1}} = \tan\frac{\theta_{2i-1}}{2}; &\frac{\psi_{2i+1}}{\psi_{2i}} = \cot\frac{\theta_{2i}}{2}, \,\,\, \,\, \forall i \in {\text{positive int}}.
\end{align}
As for the $\pi$-mode, setting $\lambda = -1$, the first column of $K$ gives
\begin{align}
    \cos\theta_1\psi_1 + \sin\theta_1\psi_2 = -\psi_1.
\end{align}
The solution is
\begin{align}
    \frac{\psi_2}{\psi_1} = \frac{-1-\cos\theta_1}{\sin\theta_1} = -\cot\frac{\theta_1}{2}.
\end{align}
The second column of $K$ gives
\begin{align}
    -\sin\theta_1\cos\theta_2 \psi_1 + \cos\theta_1\cos\theta_2 \psi_2 + \sin\theta_2 \psi_3 = -\psi_2.
\end{align}
With $\psi_2/\psi_1 = -\cot(\theta_1/2)$, one obtains
\begin{align}
    \frac{\psi_3}{\psi_2} = \frac{-1-\cos\theta_2}{\sin\theta_2} = -\cot\frac{\theta_2}{2}.
\end{align}
This pattern persists for all columns. The analytic expression of coefficients of the $\pi$-mode is
\begin{align}
    \frac{\psi_{i+1}}{\psi_i} = -\cot\frac{\theta_i}{2},\,\,\,  \forall i \in \text{positive int}.
\end{align}
Note that this analytic expression for localized edge modes is valid when $\{ \theta_i \}$ are inside the phase allowing the existence of edge modes because in this case the series is convergent in $i$. Otherwise, one has to truncate at some $\theta_M$ and obtain an approximate edge mode. In ITFIM, we identify the angles as the local transverse-field, $g_i T = \theta_{2i-1}$, and the local Ising coupling, $J_{xi}T = \theta_{2i}$ and construct a series of data points, $(g_1T, J_{x1}T), (g_2T, J_{x1}T), (g_2T, J_{x2}T), (g_3T, J_{x2}T), \ldots,$ $(g_iT,J_{xi}T),(g_{i+1}T, J_{xi}T) , \ldots$. These points form a trajectory in phase space, e.g., see Fig.~\ref{Fig: Phase diagram}. 

We now explain the condition for truncation. Starting from $(g_1T, J_{x1}T)$, one follows the trajectory in phase space until the data point falls outside the topological phase of the starting point. For example, if  $(g_1T, J_{x1}T)$ is in the $0$-mode phase, one follows the data points until one encounters a $(g_iT, J_{xi}T)$ that is inside the $0$-mode phase but a $(g_{i+1}T, J_{xi}T)$ that is outside. The data is truncated at $(g_iT, J_{xi}T)$ and we set $J_{xi}T = \theta_M = \pi/2$. In contrast, if one encounters a $(g_{i+1}T, J_{xi}T)$ that is inside the $0$-mode phase but $(g_{i+1}T, J_{xi+1}T)$ that  is outside, one truncates at $(g_{i+1}T, J_{xi}T)$ and sets $g_{i+1}T = \theta_M = \pi/2$. Then one constructs the approximate $M \times M$ matrix $\Tilde{K}_{\text{ITFIM}}$ and the approximate $0$-mode, $\Tilde{\psi}_0$, or approximate $\pi$-mode, $\Tilde{\psi}_\pi$ from the truncated Krylov angles: $\{ \theta_0, \theta_1, \ldots, \theta_M \}$, where $\theta_0 =0$ and $\theta_M = \pi/2$. 
Noting that the autocorrelation function is the return amplitude to the first site of the Krylov chain, one arrives at the approximate autocorrelation function 
\begin{align}\label{aproxsupp}
    &\Tilde{A}_\infty^\alpha(n) = |\Tilde{\psi}_{\alpha,1}|^2 e^{-\Gamma_\alpha n };
    &\Gamma_\alpha = \lim_{m \rightarrow \infty} \frac{-\ln[|(\Tilde{\psi}_{\alpha}|\Tilde{K}_{\text{ITFIM}}^m|\Tilde{\psi}_{\alpha})|]}{m},
\end{align}
where the $\alpha = 0,\pi$ label denotes the approximate $0$ or $\pi$-mode $|\Tilde{\psi}_{\alpha})$, and $\Tilde{\psi}_{\alpha,1}$ is the first element of this vector. The decay rate $\Gamma_\alpha$ is obtained from the approximation $\exp(-\Gamma_\alpha m) \approx |(\Tilde{\psi}_{\alpha}|\Tilde{K}_{\text{ITFIM}}^m|\Tilde{\psi}_{\alpha})|$. We set $m=1000$ in the numerical computations. 
\end{widetext}

\begin{figure*}
    \includegraphics[width=0.24\textwidth]{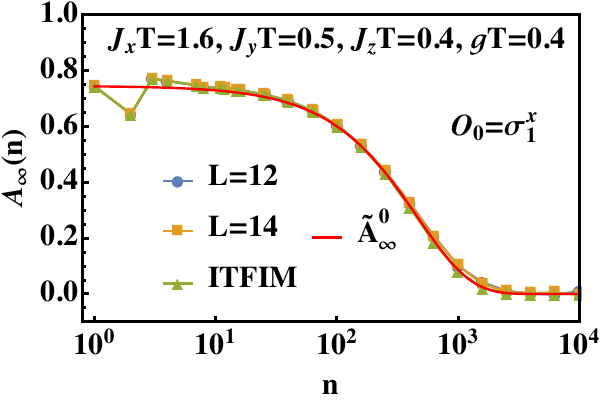}
    \includegraphics[width=0.24\textwidth]{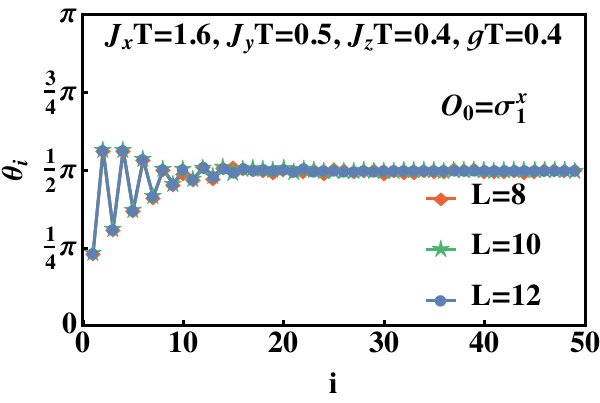}
    \includegraphics[width=0.24\textwidth]{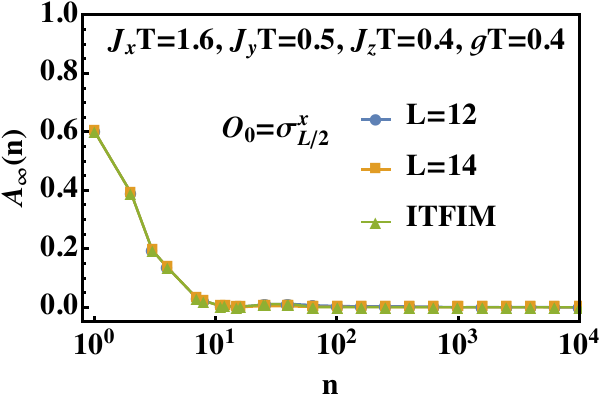}
    \includegraphics[width=0.24\textwidth]{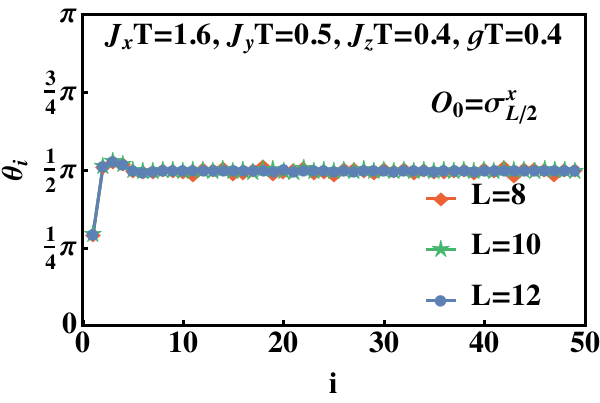}

    \caption{The infinite temperature autocorrelation function and Krylov angles of the Floquet transverse-field XYZ model \eqref{Eq: TXYZ}, generated from the edge spin $\sigma_1^x$  (two left panels) and bulk spin $\sigma_{L/2}^x$  (two right panels). $K_{\text{ITFIM}}$ is truncated to a $50 \times 50$ matrix and constructed from the Krylov angles for $L=12$. The corresponding  autocorrelation  obtained as a return amplitude, (green lines in first and third panels), and from further approximations \eqref{aproxsupp}, (red line in the first panel) show good agreement with ED. The consistency of the autocorrelation function between the Floquet transverse-field XYZ model and the ITFIM, supports the idea that any Hermitian operator in any Floquet model can be mapped to an edge Majorana in a suitable ITFIM.}
    \label{Fig: TXYZ}
\end{figure*}

\section{Dynamics of the edge and a bulk operator of the Floquet Transverse-field XYZ model}
\label{Sec:D}
In this section, we explore the mapping of the  Floquet transverse-field XYZ model (TXYZ) to the ITFIM, by studying the dynamics of an edge operator as well as a bulk operator of the TXYZ. The unitary evolution is given by  
\begin{align}
    U_\text{TXYZ} = e^{-i\frac{T}{2}g H_z}e^{-i\frac{T}{2}J_z H_{zz}}e^{-i\frac{T}{2}J_y H_{yy}}e^{-i\frac{T}{2}J_x H_{xx}},\label{Eq: TXYZ}
\end{align}
where
\begin{align}
    &H_{xx} = \sum_{i=1}^{L-1} \sigma_i^x\sigma_{i+1}^x; &H_{yy} = \sum_{i=1}^{L-1} \sigma_i^y\sigma_{i+1}^y;\nonumber\\
    &H_{zz} = \sum_{i=1}^{L-1} \sigma_i^z\sigma_{i+1}^z; &H_{z} = \sum_{i=1}^{L} \sigma_i^z.
\end{align}

\begin{figure}[h!]
    \centering
    \includegraphics[width=0.3\textwidth]{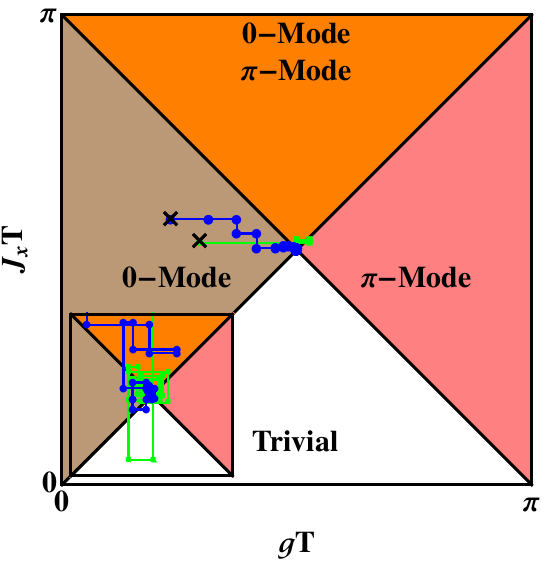}
    \caption{The trajectory of local couplings from the edge to the bulk for the ITFIM, where $\theta_{2j-1}$ ($\theta_{2j}$) corresponds to the local transverse-field $g_jT$ (Ising coupling $J_{xj}T$). The Krylov angles are for $L=12$ in Fig.~\ref{Fig: TXYZ} with the black crosses
    denoting the local couplings on the first Krylov site. There are no triangles unlike Figure \ref{Fig: Phase diagram} since this model cannot be connected to the TFIM by turning off the transverse-field. The two trajectories correspond to Krylov angles from different initial operators, $\sigma_1^x$ (blue circles) and $\sigma_{L/2}^x$ (green squares). The first few blue circles are within the $0$-mode phase but the green squares jump between different phases immediately.  This reflects that the lifetime of $\sigma_1^x$ is much longer than $\sigma_{L/2}^x$. Eventually both trajectories approach the center of the phase diagram.}
    \label{Fig: Phase diagram 2}
\end{figure}

We consider two setups corresponding to the seed operator being the edge (bulk) operator, $O_0 = \sigma_1^x$ ($O_0 = \sigma_{L/2}^x$). We take the couplings to be $J_xT = 1.6$, $J_yT = 0.5$, $J_zT = 0.4$ and $gT = 0.4$. Thus one may regard this as a Floquet XYZ model perturbed by a transverse field $gT$. Note that the TFIM is not the non-perturbed limit of this model. Nevertheless, its autocorrelation function can be well described by the ITFIM as shown in Fig.~\ref{Fig: TXYZ}. The different lifetimes of the edge and bulk operators is reflected in the difference between the trajectory of the local couplings in the phase diagram Fig.~\ref{Fig: Phase diagram 2}. The first few blue circles ($O_0 = \sigma_1^x$) are within the $0$-mode phase but the green squares ($O_0 = \sigma_{L/2}^x$) leave the $0$-mode phase immediately. 

\begin{figure*}
    \centering
    \includegraphics[width=0.3\textwidth]{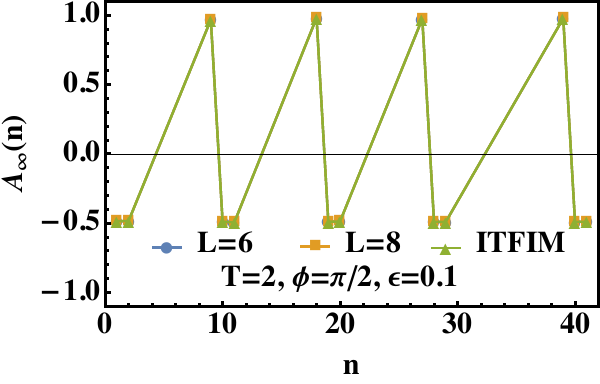}
    \includegraphics[width=0.29\textwidth]{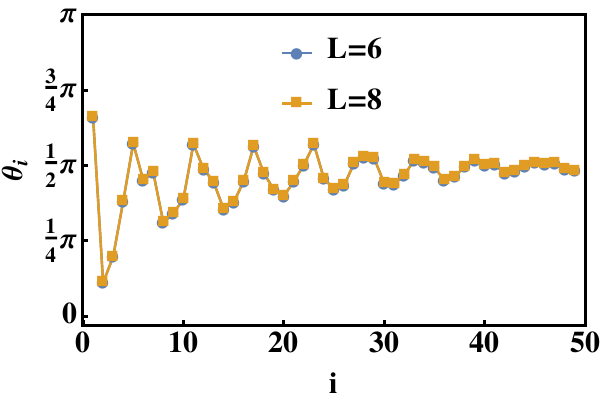}
    \includegraphics[width=0.19\textwidth]{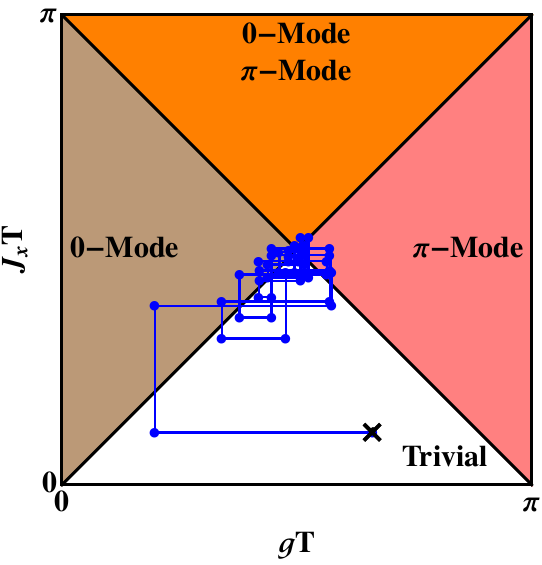}
    \includegraphics[width=0.3\textwidth]{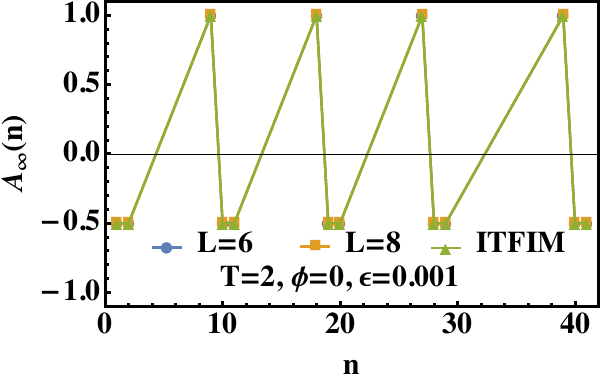}
    \includegraphics[width=0.29\textwidth]{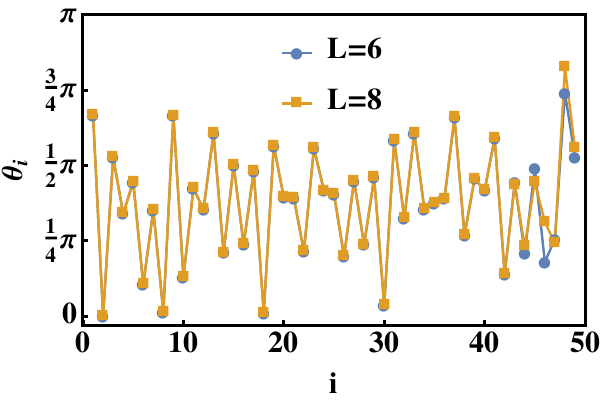}
    \includegraphics[width=0.19\textwidth]{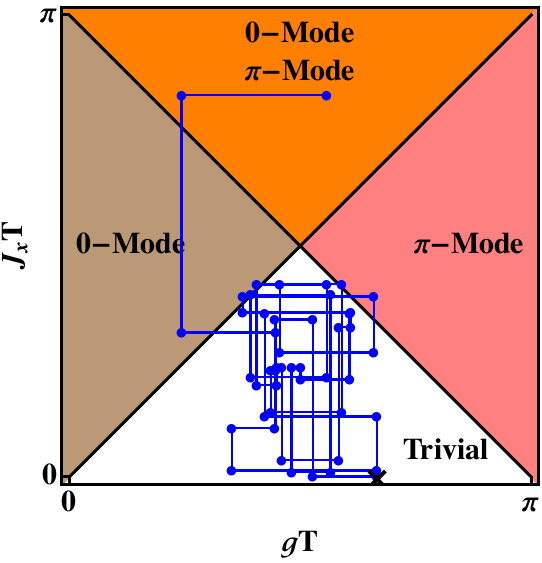}
    
    \caption{Same as Fig.~\ref{Fig: Z3 clock model}, but now the autocorrelation functions of the $Z_3$ clock model are plotted on the linear time scale (left panels). This scale shows the period tripling more clearly. The corresponding unfolded Krylov angles from the unitary (middle panels) and their trajectories in the phase diagram of the TFIM (right panels) are also shown.}
    \label{Fig: Supp Linear scale}
\end{figure*}

\section{Period tripled autocorrelation function on a Linear Scale with unfolded Krylov angles}
\label{Sec:E}
We present the autocorrelation functions of the $Z_3$ clock model on the linear time scale in Fig.~\ref{Fig: Supp Linear scale}. The corresponding unfolded Krylov angles obtained from $U_2$, and their trajectories in the phase diagram are also shown. In contrast, Fig.~\ref{Fig: Z3 clock model} in the main text shows the folded Krylov angles obtained from $U_2^3$

\section{System size effects of the ITFIM on the $Z_3$ clock model}
\label{Sec:F}
\begin{figure}[h!]
    \centering
    \includegraphics[width=0.4\textwidth]{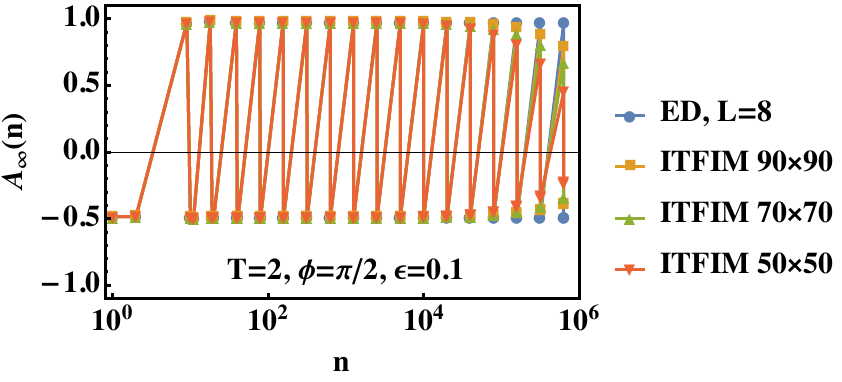}    
    \includegraphics[width=0.4\textwidth]{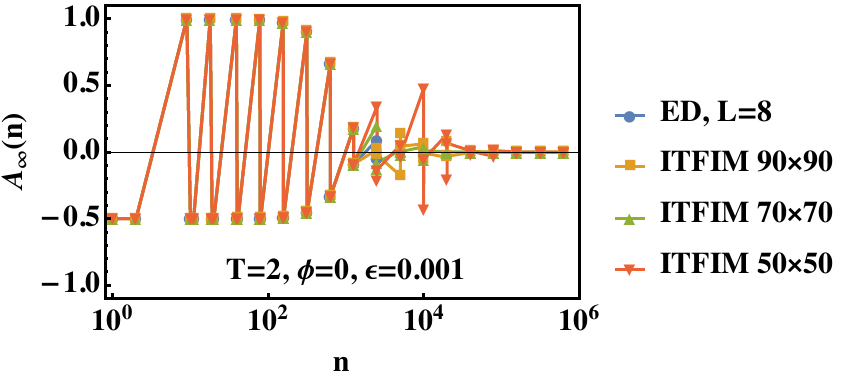}
    \caption{Infinite temperature autocorrelation functions  for the $Z_3$ clock model, and corresponding to the two left panels of Fig.~\ref{Fig: Z3 clock model}. ITFIM results with different sizes of truncation, $50\times 50$, $70\times 70$ and $90\times 90$, are plotted on top of the ED results. Increasing the system size of the ITFIM improves the agreement with ED, but the late-time oscillation on the right panel are not removed. The zoom-in of the latter is shown in the top panel of Fig.~\ref{Fig: Supp Z3 zoom-in/log-log}.} 
    \label{Fig: Supp Z3}
\end{figure}

\begin{figure}[h!]
    \centering
    \includegraphics[width=0.4\textwidth]{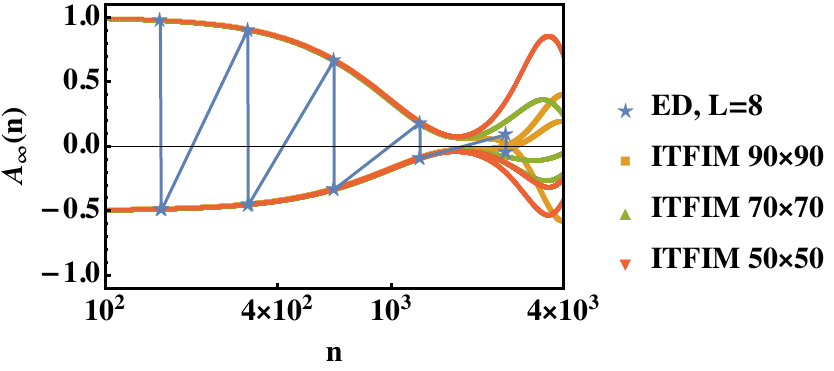}
    \includegraphics[width=0.3\textwidth]{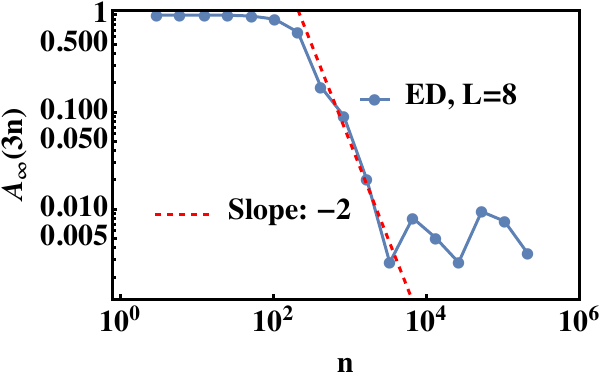}
    \caption{The zoomed-in plot (top panel) and the period folded Log-Log plot (bottom panel) of the autocorrelation function in the bottom panel of Fig.~\ref{Fig: Supp Z3}. Top panel: increasing system size of the ITFIM improves the agreement with ED. The late-time oscillations and breaking of period tripling due to the truncation are postponed to later times for larger system sizes. Bottom panel: the period folded autocorrelation follows a power-law decay $\sim 1/n^2$ in time.}
    \label{Fig: Supp Z3 zoom-in/log-log}
\end{figure}

In this section we present a more detailed discussion of the $Z_3$ clock model. The ITFIM for different system sizes, $50\times 50$, $70\times 70$ and $90\times 90$, are shown in Fig.~\ref{Fig: Supp Z3}. For the long-lived period tripled edge mode (top panel), increasing the truncated system size of the ITFIM improves the agreement with ED at late times. 
The effect of increasing system size for the short-lived autocorrelator are shown in the bottom panel of Fig.~\ref{Fig: Supp Z3}, with its zoomed in version shown in the top panel of Fig.~\ref{Fig: Supp Z3 zoom-in/log-log}. The truncation leads to oscillations and a strong breaking of period tripling. 
By increasing the system size, these effects are pushed to later times, but do not disappear.
The hard truncation by setting the last Krylov angle to $\pi/2$ cannot faithfully reproduce the bulk property in the thermodynamic limit, which is shown by the inconsistency at late times in the top panel of Fig.~\ref{Fig: Supp Z3 zoom-in/log-log}. Moreover, if we period fold the autocorrelation by considering $A_{\infty}(3n)$ for $n$ being integer, it shows a power-law decay in the Log-Log scale (bottom panel of Fig.~$\ref{Fig: Supp Z3 zoom-in/log-log}$). This shows that the short-lived correlator requires 
a different effective model than \eqref{Eq: Approx} to describe its dynamics, and in particular one which is more sensitive to the details of the bulk Krylov angles.


%

\end{document}